\documentclass{article}

\usepackage[authoryear,longnamesfirst]{natbib}
\usepackage{PRIMEarxiv}
\usepackage{amsmath}
\usepackage[utf8]{inputenc} 
\usepackage[T1]{fontenc}    
\usepackage{hyperref}       
\usepackage{url}            
\usepackage{booktabs}       
\usepackage{amsfonts}       
\usepackage{nicefrac}       
\usepackage{microtype}      
\usepackage{lipsum}
\usepackage{fancyhdr}       
\usepackage{graphicx}       
\graphicspath{{media/}}     
\usepackage[toc,page]{appendix}
\DeclareMathOperator{\diag}{diag}
\DeclareMathOperator{\Tr}{Tr}

\title{EM algorithm for generalized Ridge regression with spatial covariates} 

\author{
  Said Obakrim\\
  IRMAR \\
  Université de Rennes 1\\
  \texttt{said.obakrim@univ-rennes1.fr} \\
    \And
  Pierre Ailliot\\
  LMBA\\
  Université de Bretagne Occidentale\\
  \texttt{pierre.ailliot@univ-brest.fr} 
   \And
  Valérie Monbet \\
  IRMAR \\
  Université de Rennes 1\\
  \texttt{valerie.monbet@univ-rennes1.fr} 
  \And
  Nicolas Raillard\\
  LCSM\\
  Ifremer\\
  \texttt{nicolas.raillard@ifremer.fr} 
}

\begin{document}

\maketitle

\begin{abstract}
The generalized Ridge penalty is a powerful tool for dealing with overfitting and for high-dimensional regressions. The generalized Ridge regression can be derived as the mean of a posterior distribution with a Normal prior and a given covariance matrix. The covariance matrix controls the structure of the coefficients, which depends on the particular application. For example, it is appropriate to assume that the coefficients have a spatial structure in spatial applications. This study proposes an expectation-maximization algorithm for estimating generalized Ridge parameters whose covariance structure depends on specific parameters. We focus on three cases: diagonal (when the covariance matrix is diagonal with constant elements), Matérn, and conditional autoregressive covariances. A simulation study is conducted to evaluate the performance of the proposed method, and then the method is applied to predict ocean wave heights using wind conditions.   
\end{abstract}
\section{Introduction}
Consider an experiment where we have the data $\{y,X\}$, of $n$ observations of a continuous variable $Y$ and $n\times d$ matrix of covariates X. Suppose that $Y$ is related to $X$ via a linear model 
\begin{equation}
   Y = X\beta + \epsilon,
   \label{eq:1}
\end{equation}
where $\beta$ are model coefficients and $\epsilon \sim \mathcal{N}(0,\sigma^2)$ is the model error. We suppose that the intercept is either included in $\beta$ (so that the first column of X is a vector of 1) or that $Y$ and $X$ are centered.
The least squares estimates are the best linear unbiased estimates of the parameters $\beta$. However, in the case of multicollinearity or high-dimensionality, penalized linear regression methods, like Ridge regression, are needed to control the variance. Ridge estimator of the problem  \eqref{eq:1} is
\begin{equation}
    \hat{\beta}_\lambda^{Ridge} = \arg \min_{\beta} -\ell(\beta,\sigma^2) + \lambda \lVert\beta\rVert^2 
    \label{eq:2}
\end{equation}
where $\lambda$ is the regularization parameter and $\ell(\beta,\sigma^2)$ is the log-likelihood of the model \eqref{eq:1}. High values of $\lambda$ permit to reduce the variance and increase the bias of the model. A good model should have a trade-off between variance and bias \citep{hastie2009elements}. In order to find a trade-off between bias and variance, the hyperparameter $\lambda$ needs to be selected.

\cite{boonstra2015small} classified methods for selecting $\lambda$ into goodness-of-fit-based and likelihood-based methods. Goodness-of-fit-based methods define a goodness of fit criterion (such as the mean squared error) and minimize it in terms of $\lambda$. The most common goodness-of-fit-based method is the k-fold cross-validation which consists of partitioning observations into k groups and estimating $\beta$ k times for each $\lambda$ leaving out one group. For each $\lambda$, a goodness of fit score is calculated, and $\lambda$ with the maximum score value is chosen. The typical choice of k is 5 and 10, while setting $k=n$ leads to leave-one-out cross-validation (LOOCV). LOOCV leads to a better estimation of $\lambda$; however, it is computationally expensive given that it requires fitting the model $n$ times \citep{patil2021uniform}. Generalized cross-validation (GCV) \citep{gcv} is an approximation of LOOCV that does not require fitting n models. GCV uses a weighted version of the predicted residual error sum of squares (PRESS) statistic \citep{allen1974relationship} as a goodness of fit criterion. One of the problems with goodness-of-fit-based methods is the selection of the grid of $\lambda$, which influences the estimation. 

Assuming that $Y|\beta \sim \mathcal{N}(X\beta,\sigma^2I_n)$, Ridge regression can be derived as the mean of a posterior distribution with the prior $\beta \sim \mathcal{N}(0_d,\sigma^2\lambda^{-1}I_d)$ \citep{Ridge} and as in Bayesian hierarchical linear regression, likelihood-based methods maximize the likelihood with respect to $\sigma^2$ and $\lambda$ using for instance an iterative method \citep{lee1996hierarchical}. Unlike goodness-of-fit-based methods, the advantage of likelihood-based approaches is, on the one hand, that they do not require grid selection for the regularization parameters. On the other hand, likelihood-based methods can be generalized to consider any form of prior for the coefficients $\beta$. In some applications, the regression coefficients can be penalized differently, or a joint penalization of the coefficients is required. For example, in spatial statistics, where predictors have a spatial structure, it is reasonable to suppose that coefficients have a spatial structure. To do that, the generalized Ridge \citep{Ridge} can be used. Generalized Ridge extends the equation \eqref{eq:2} by replacing the term $\lambda \lVert\beta\rVert^2$ to $\beta^T\Delta\beta$, where $\Delta$ is called the penalty matrix. In general, $\Delta$ depends on some regularization parameters (see, e.g., \cite{goeman2008autocorrelated} and \cite{hemmerle1975explicit}); however, when the number of the regularization parameters is greater than 1, goodness-of-fit-based methods struggle with the problem of combinatorial explosion. Generalized Ridge in the hierarchical linear model framework, is equivalent to suppose that $\beta \sim \mathcal{N}(0_d,\Sigma_\theta)$ where $\Sigma_\theta$ is a covariance matrix that depends on some parameters $\theta$. Note that $\Sigma_\theta$ corresponds to the inverse of the penalty matrix $\Delta$. The classical Ridge is a special case of this model when the covariance matrix $\Sigma_\theta$ is diagonal, and $\theta$ is the usual regularization parameter $\lambda$.

Considering $\beta$ as a hidden variable, \cite{bishop} proposed an expectation-maximization (EM) algorithm to find the maximum likelihood estimation (MLE) of parameters of a Bayesian linear regression model. The EM algorithm \citep{em} is a method for estimating the parameters of a model with hidden variables. The EM algorithm alternates between two steps: the expectation and maximization steps. The E-step calculates the conditional expectation of the log-likelihood given the observations and current parameters. In the M-step, the parameters are estimated by maximizing the conditional expectation of the log-likelihood calculated in the E-step. In this study, we extend the algorithm in \cite{bishop} and propose an EM algorithm to estimate the parameters of hierarchical linear regression when $\beta \sim \mathcal{N}(0,\Sigma_\theta)$. At first, we study the case where $\Sigma_\theta$ is diagonal with constant elements, which corresponds to the classical Ridge in equation \eqref{eq:2} and the problem studied by \citep{bishop}. Then, we consider the case where the coefficients $\beta$ have a spatial structure, especially when $\Sigma_\theta$ is the Matérn or the conditional autoregressive (CAR) covariance. A simulation study is done to assess the performance of the method. Then, the proposed method is applied to oceanographic data where the response variable represents a wave parameter in a location in the Bay of Biscay, and $X$ represents wind conditions over the North Atlantic \citep{statistical}.

This paper is organized as follows. The proposed method and its special cases are presented in Section 2. Then, a simulation study is conducted in Section 3 to assess the performance of the proposed method. In section 4, we apply the methodology to oceanography data. Finally, this study is concluded in Section 5. 

\section{Proposed method}
As stated in the introduction, Ridge regression can be viewed as a hierarchical linear model  where $\beta \sim \mathcal{N}(0_d,\sigma^2\lambda^{-1}I_d)$. When there is a structure on the coefficients, it is unreasonable to consider all possible covariance functions as possible candidates for $\beta$. Therefore, we suppose that the covariance of $\beta$ depends on some parameters $\theta$, so that $\beta \sim \mathcal{N}(0_d,\Sigma_\theta)$. This motivates using the EM algorithm to find the maximum likelihood estimation of the parameters, where the model parameters are then $\Theta = (\sigma^2, \theta)$. The proposed method is described in this section, and three special cases of the covariance $\Sigma_\theta$ (the diagonal, Matérn, and CAR) are studied.  
 
\subsection{EM algorithm for generalized Ridge}
Consider the linear model \eqref{eq:1} and assume that $\beta$ is a latent variable that follows a normal distribution. We define the regression model hierarchically as  
\begin{equation}
      \begin{split}
            &\beta \sim \mathcal{N}(0_d,\Sigma_\theta)\\
            &Y \mid \beta, \Theta  \sim \mathcal{N}(X\beta, \sigma^2I_n)
      \end{split}
    \label{eq:3}
\end{equation}
where $\Theta = (\sigma^2, \theta)$. Note that for simplicity, we assume that the mean of $\beta$ is zero. The EM algorithm for the case where $\beta$ has a non-zero mean will be presented in the Appendix.

Given a sample $y = (y_1,...,y_n)$,
the complete log-likelihood is expressed as 
\begin{equation}
\begin{split}
        \ln p(y, \beta;\Theta) &= \ln p(y \mid \beta; \sigma^2)+ \ln p(\beta; \theta) \\
        &=  -\frac{1}{2}\left(d\ln(2\pi)+\ln(|\Sigma_\theta|)+\beta^T\Sigma_\theta^{-1}\beta + n\ln(2\pi) +n\ln(\sigma^2) + \frac{1}{\sigma^2}\lVert y - X\beta \rVert^2\right)
\end{split}
\label{eq:4}
\end{equation}
Maximum likelihood estimation consists of maximizing \eqref{eq:4} with respect to the parameters $\Theta$. This is usually done with the Expectation-Maximization algorithm in the latent variable context. The EM algorithm alternates between the E-step and M-step. In the E-step, the expectation $Q(\Theta |\Theta ^{(t)})$ of the complete likelihood with respect to the posterior distribution of the latent variable $\beta$ and the parameters $\Theta^{(t)}$ from the previous iteration $t$ is calculated. In the M-step, the quantity $Q(\Theta|\Theta^{(t)})$ is maximized with respect to the parameters $\Theta$.

The E-step and M-step are defined as follows
\begin{itemize}
    \item E-step: \\
    \begin{equation}
        Q(\Theta
        |\Theta^{(t)}) = \mathbb{E}(\ln p(y,\beta;\Theta)\mid y,\Theta^{(t)}).
        \label{eq:5}
    \end{equation}
\end{itemize}
The posterior distribution of the latent variable $\beta$ is a normal distribution with mean $\mu_{\beta \mid y}$ and covariance matrix $\Sigma_{\beta \mid y}$ such that 
\begin{equation}
\left\{
    \begin{array}{ll}
        \Sigma_{\beta|y} = 
        (\Sigma_\theta^{-1} + \frac{1}{\sigma^2}X^TX)^{-1}\\
        \mu_{\beta|y} = (X^TX +\sigma^2\Sigma_{\theta}^{-1})^{-1}X^Ty.
    \end{array}
\right.
\label{eq:6}
\end{equation}
Note that $\mu_{\beta|y}$ defined in (\ref{eq:6}) is a generalized Ridge estimator (see e.g. \cite{Ridge}) solution of the optimization problem 
\begin{equation} 
\label{eq:gRidge}
    \mu_{\beta|y} = \arg \min_{\beta} \frac{ \lVert y-X\beta \rVert^2}{\sigma^2} + \beta^T  \Sigma_{\theta}^{-1} \beta
\end{equation}
Therefore, 
\begin{equation}
     Q(\Theta|\Theta^{(t)}) =  -\frac{1}{2}\left(\ln(|\Sigma_\theta|) + \mathrm{Tr}(\Sigma_\theta^{-1}\mathbb{E}(\beta\beta^T\mid y,\Theta^{(t)}))+\ln(\sigma^2)+\frac{1}{\sigma^2}\mathbb{E}(\lVert y-X\beta\rVert^2\mid y,\Theta^{(t)})\right ) + C
     \label{eq:7}
\end{equation}
where $C$ is a constant and 
\begin{equation}
    \left\{
    \begin{array}{ll}
        \mathbb{E}(\beta\beta^T|y;\Theta^{(t)}) = \Sigma_{\beta|y} + \mu_{\beta|y}\mu_{\beta|y}^T\\
        \mathbb{E}(\lVert y - X\beta\rVert^2|y;\Theta^{(t)}) = \lVert y\rVert^2 -2y^TX\mu_{\beta|y}+
        \mathrm{Tr}(X^TX\mathbb{E}(\beta\beta^T|y;\Theta^{(t)}))
    \end{array}
\right.
\label{eq:8}
\end{equation}
\begin{itemize}
    \item M-step:
\end{itemize}
The maximization step computes
\begin{equation}
    \Theta^{(t+1)} = \arg \max_{\Theta}  Q(\Theta|\Theta^{(t)})
    \label{eq:9}
\end{equation}
which leads to the following updates of the parameters $\sigma^2$ and $\theta$
\begin{equation}
\begin{split}
    &\sigma^{2,(t+1)} = \frac{1}{n}(\lVert y\rVert^2 -2y^TX\mu_{\beta|y}+
       \mathrm{Tr}(X^TX\mathbb{E}(\beta\beta^T|y;\Theta^{(t)})))\\
    & \theta^{(t+1)} = \arg \max_{\theta} \, \, \ln(|\Sigma_\theta^{-1}|) - \mathrm{Tr}(\Sigma_\theta^{-1}\mathbb{E}(\beta\beta^T\mid y,\Theta^{(t)}))
\end{split}
\label{eq:10}
\end{equation}

\subsection{Special cases}
The M-step in equation \eqref{eq:10} requires the maximization of $Q(\Theta|\Theta^{(t)})$ over the parameters of the covariance $\Sigma_\theta$. In this study, we will explore three cases. First, we consider the case where $\Sigma_\theta$ is diagonal. Then, the case where $\beta$ has a spatial structure, especially when the parametric covariance is the Matérn covariance function. Finally, we consider the conditional autoregressive model (CAR).
\subsubsection{Diagonal case}
In the classical Ridge, the covariance matrix of the coefficients $\beta$ is supposed to be diagonal such that 
\begin{equation}
    \Sigma_\theta = \sigma_\beta^2\mathbf{I_d}.
    \label{eq:11}
\end{equation}
The M-step of the covariance in \eqref{eq:10} becomes 
\begin{equation}
    \sigma_\beta^{2,(t+1)} = \arg \max_{\sigma_\beta^2} \, \, -d\ln(\sigma_\beta^2) - \frac{1}{\sigma_\beta^2}\mathrm{Tr}(\mathbb{E}(\beta\beta^T\mid y,\Theta^{(t)})).
\label{eq:12}
\end{equation}
Setting the derivatives with respect to $\sigma^2_\beta$ to zero, we obtain the M-step 
\begin{equation}
    \sigma_\beta^{2,(t+1)} =\frac{\mathrm{Tr}(\mathbb{E}(\beta\beta^T\mid y,\Theta^{(t)}))}{d}.
\label{eq:13}
\end{equation}

Note that $\frac{1}{\sigma^2_\beta}$ corresponds to the regularization parameter $\lambda$ in equation \eqref{eq:1}. As stated in the introduction, Ridge regression requires the selection of the regularization parameter. Therefore, the EM algorithm can be an alternative to cross-validation for estimating Ridge coefficients along with the regularization parameter. A comparison of the two methods (cross-validation and EM algorithm) is given in the Appendix. 
\subsubsection{Spatial covariance functions}
In spatial statistics applications, one may assume that $\beta$ has a spatial structure. One way to do that is to assume that $\beta$ has a parametric covariance function. There are many choices of covariance functions that are widely used for Gaussian processes and kriging \citep{gp}.
In this study, we focus on the stationary Matérn covariance, which has the form
\begin{equation}
    K(h;\phi,\kappa) =\frac{\sigma^2_\beta}{2^{\kappa-1} \Gamma(\kappa)} \left(\frac{h}{\phi}\right)^\kappa K_\kappa\left(\frac{h}{\phi}\right)
\end{equation}
where $h$ is the distance between two points, $\Gamma$ is the Gamma function, and $K_\kappa$ is the modified Bessel function \citep{handbook}. The Matérn function is parameterized by the variance parameter $\sigma_\beta^2$, the range parameter $\phi$, and the smoothness parameter $\kappa$. The range parameter $\phi$ controls the decay rate with distance, with larger values of $\phi$ corresponding to more strongly correlated variables, and the smoothness parameter $\kappa$ controls the mean-square differentiability of the spatial process. 

The M-step of the covariance of $\beta$ in \eqref{eq:11} becomes
\begin{equation}
    (\sigma_\beta^{2,(t+1)},\theta^{(t+1)}) = \arg \max_{\sigma^2_\beta,\theta} \,\, \ln(| R_\theta^{-1}|)- d\ln(\sigma^2_\beta) - \frac{1}{\sigma^2_\beta}\mathrm{Tr}( R_\theta^{-1} \mathbb{E}(\beta\beta^T\mid y,\Phi^{(t)}))
    \label{eq:eq15}
\end{equation}
where $R_\theta$ is the Matérn correlation and $\theta = (\phi, \kappa)$. Since the variance parameter is constant and following \cite{bachoc}, the optimization of the variance parameter $\sigma^2_\beta$ can be carried out separately with the correlation parameters $\phi$ and $\kappa$. Therefore, 
\begin{equation}
    \begin{split}
       &\sigma_\beta^{2,(t+1)} = \frac{\mathrm{Tr}( R_\theta^{-1} \mathbb{E}(\beta\beta^T\mid y,\Phi^{(t)}))}{d}\\
       &\theta^{(t+1)} = \arg \max_{\theta}\,\, \ln(|R_\theta^{-1}|) - d\ln(\mathrm{Tr}( R_\theta^{-1} \mathbb{E}(\beta\beta^T\mid y,\Phi^{(t)}))).
    \end{split}
    \label{eq:16}
\end{equation}
The solution to the optimization problem in equation \eqref{eq:16} cannot be done analytically; therefore, numerical optimization algorithms are used. This study uses the quasi-Newton method L-BFGS-B to optimize the parameters. Given the difficulties in estimating Matérn parameters \citep{kaufman2013role}, we a priori fix the smoothness parameter as $\frac{3}{2}$, which gives the classical $\frac{3}{2}$-Matérn covariance function.   
\subsubsection{Conditional autoregressive model}
The M-step in equation \eqref{eq:9} requires the inversion of the covariance matrix, which can be challenging for large matrices. This problem is wildly discussed in Gaussian processes literature \citep{ambikasaran2015fast, storkey1999truncated}. Therefore, it can be numerically advantageous to parameterize the precision matrix (inverse of the covariance matrix) instead of the covariance matrix. This is motivated by the fact that the precision matrix $P_\theta = \Sigma_\theta^{-1}$ can be approximated by a sparse matrix \citep{tajbakhsh2020theoretical}. In fact, the off-diagonal elements of the precision matrix correspond to the conditional covariance between two variables given the remaining variables. Therefore, conditionally independent variables have zero values in the precision matrix. 

Gaussian Markov random fields (GMFs) are wildly used in spatial statistics \citep{cressie2015statistics}. GMFs models have a Markov property making them computationally and theoretically suitable \citep{rue2001fast}. Furthermore, \citep{rue2002fitting} demonstrated that a GMF model can approximate a Gaussian field with a Matérn correlation function and other families of correlation functions. Conditional autoregressive (CAR) models are classes of GMFs with well-defined joint Gaussian distribution \citep{cressie2008some}. This subsection will study cases where the coefficients $\beta$ have the CAR model property. The joint distribution of a CAR is expressed as
\begin{equation}
    \beta \sim \mathcal{N}(0,\tau^{2}(I_d - \alpha H)^{-1} \Phi).
    \label{eq:17}
\end{equation}
The distribution of $\beta$ depends on unknown parameters $\alpha$ and $\tau^2$, and many types of CAR models depend on the choice of the matrix $H$  and $\Phi$. Following \citep{besag1991bayesian}, in this study, we consider the Weighted CAR (WCAR) model where 
\begin{equation}
    \Phi = \diag(\lvert N_1 \rvert^{-1},..., \lvert N_d \rvert^{-1})
    \label{eq:18}
\end{equation}
where $\lvert N_i \rvert$ is the number of neighbors of location $i$ and $H = \left(\frac{a_{ij}}{\lvert N_i \rvert}\right)_{d\times d}$; $i,j = 1,..., d$, where $a_{ij}$ is the $(i,j)$ element of the adjacency matrix $A = (a_{ij})_{d\times d}$, where $a_{ij} = a_{ji} = 1$ if and only if location $i$ and $j$ are neighbors and otherwise $a_{ij} = 0$. Putting $P_\theta = \tau^{-2}(I_d - \alpha H) \Phi^{-1}$, the second part of the M-step in the equation \eqref{eq:10} becomes 
 \begin{equation}
      \theta^{(t+1)} = \arg\max_{\theta} \, \, \ln(|P_\theta|) - \Tr(P_\theta
      \mathbb{E}(\beta \beta^T\mid y,\Phi^{(t)}))
     \label{eq:19}
 \end{equation}
 where $\theta = (\tau^2, \alpha) $. 
 
As for the Matérn covariance, the solution to the optimization problem \eqref{eq:19} cannot be done analytically, and the numerical optimization algorithm L-BFGS-B is used. Note that the optimization of the variance parameter $\tau^2$ can also be carried out separately with the parameter $\alpha$.
 
 Remark that this leads to a spatial extension of the fused Ridge method proposed in \citep{goeman2008autocorrelated}. When $\alpha=1$, we obtain 
 \begin{equation}
      \frac{1}{\tau^2}\beta^T  \Phi ^{-1}(I_d - \alpha H) \beta = \frac{1}{2\tau^2} \sum_{(i,j)|a_{ij}=1} (\beta_i-\beta_j)^2.
 \end{equation}
 This shows that any spatial coefficient variations will be penalized when solving \eqref{eq:gRidge}. In this case, replacing the L2 norm with the L1 norm leads to the fused LASSO  method proposed in \citep{tibshirani2005sparsity}. However, the matrix $(I_p - \alpha H)$ is semi-positive definite when $\alpha=1$ and thus $\Sigma_\theta$ is degenerate. Hereafter we impose the constraints $|\alpha|<1$ to ensure that the precision matrix is positive definite. Another strategy would consist of adding a regular Ridge penalty  (e.g., the discussion in \cite{Ridge}).

\section{Simulation study}
In this section, a simulation study is conducted to assess the performance of the proposed method for estimating model parameters for the three cases: diagonal, Matérn, and CAR. 
\subsection{Setup}
This study focuses on using the proposed method for spatial applications. Therefore, we consider a $15 \times 15$ regular spatial grid in a square domain $[1,15]^2$ where each location $j$  has a covariate $x_j$. We generate $X = (x_{ij})_{n\times d} $ of $n$ independent and identically distributed observations from a multivariate normal distribution with zero mean and a Matérn covariance with some arbitrary parameters $(\sigma_x^2, \phi_x, \kappa_x) = (6 ,2, 3/2)$. Then, the coefficients $\beta$, kept the same for all observations, are simulated using either the diagonal, Matèrn, or CAR case. Finally, for a given $\sigma^2$, $Y$ is simulated from the normal distribution according to equation \eqref{eq:3}. 

The parameters chosen for each case are:
\begin{itemize}
    \item Diagonal: $\sigma^2 = 36$ and $\sigma_\beta^2 = 7$ 
    \item Matérn: $\sigma^2 = 36$, $\sigma_\beta^2 = 0.1$ and $\phi = 4$
    \item CAR: $\sigma^2 = 36$, $\tau^2 = 1$ and $\alpha = 0.9$ 
\end{itemize}
The parameters are chosen so that the results of the three methods are comparable. For the CAR model, we consider four neighbors to construct the adjacency matrix, and we chose $\alpha = 0.9$ to sufficiently smooth the resulting coefficients.\\

The EM algorithm is initialized with an arbitrary set of parameters, and the E-step and M-step are repeated until no further improvement can be made to the likelihood value or to limit the computational cost until a maximum number of iterations is reached. The computation time for one iteration on an i5-7500 CPU and 16Go computer is 0.16, 3, and 1.8 seconds for diagonal, Matérn, and CAR, respectively.  
\subsection{Results}
 \begin{figure}
    \centering
    \includegraphics[width=145mm]{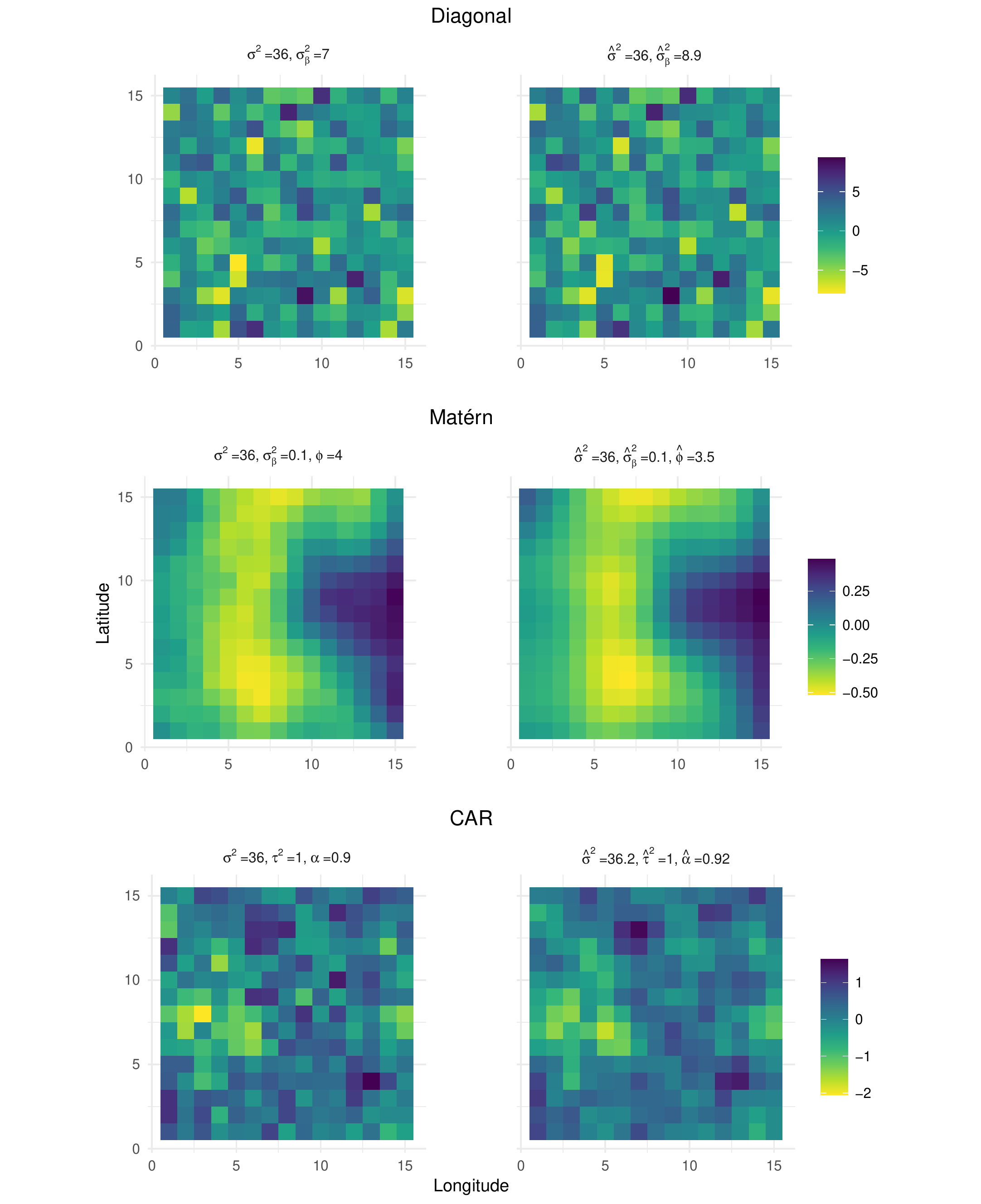}
    \caption{Simulation results for the three cases (diagonal, CAR, and Matérn). The left panels correspond to the true $\beta$ coefficients with the true parameters given in section 3.1, and the right panels correspond to the $\beta$ estimated when the sample size $n = 800$.}
    \label{fig:fig1}
\end{figure}

\begin{figure}
    \centering
    \includegraphics[width=150mm]{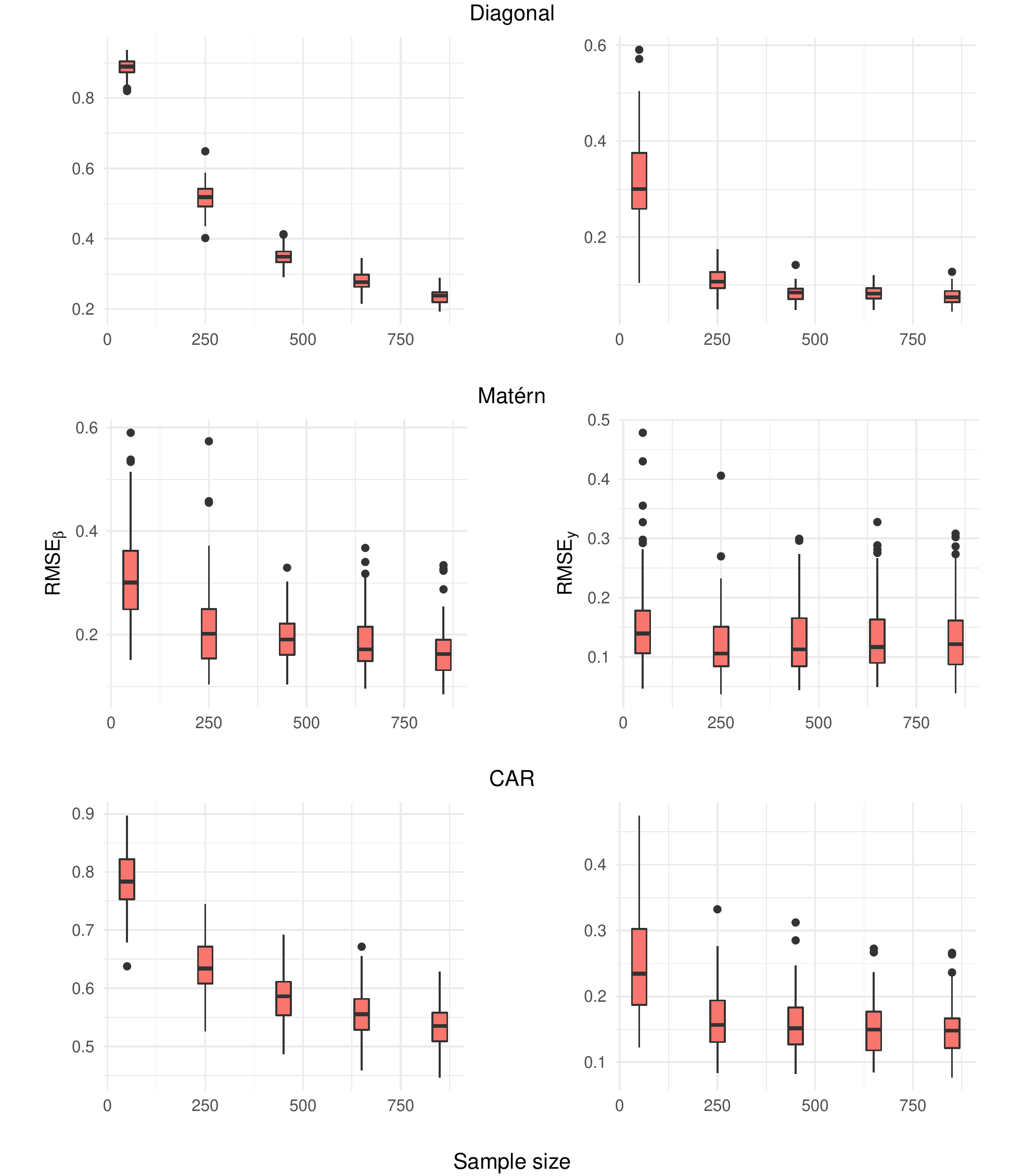}
    \caption{Results of $RMSE_\beta$ (left panels) and $RMSE_y$ (right panels) for the diagonal, CAR, and Matérn case as a function of the sample size varying from 50 to 850.}
    \label{fig:fig2}
\end{figure}

At first, one simulation is done for each case (diagonal, Matérn, and CAR) with $n=800$. The parameters are estimated using the EM algorithm presented in the previous section. Figure \ref{fig:fig1} shows the first simulation results. Left panels correspond to the true $\beta$, and right panels correspond to the estimated $\beta$ using the EM algorithm. For all the cases, the EM algorithm does well in estimating the parameters, especially the variance $\sigma^2$. 

To assess the influence of the sample size on the estimations, for each case, we perform 100 independent random simulations for each sample size varying from 50 to 850. For each simulation, the EM algorithm is used to estimate the parameters. 
Figure \ref{fig:fig2} shows the normalized root mean square error $NRMSE_\beta$ and $NRMSE_y$ for the three cases where
\begin{equation}
    \begin{split}
        &NRMSE_\beta = \frac{\sqrt{\frac{1}{d}\sum_j^d (\beta_j-\hat{\beta_j})^2}}{\hat{\sigma_\beta}}\\
        &NRMSE_y = \frac{\sqrt{\frac{1}{{n\prime}}\sum_i^{n\prime} (y_i-\hat{y_i})^2}}{\hat{\sigma}_y}
    \end{split}
\end{equation}
where $\hat{\beta_j}$ and $\hat{y_i}$ are the estimated $\beta_j$ and $y_i$ and $\hat{\sigma_\beta}$ and $\hat{\sigma}_y$ are the sample standard deviation of $\beta$ and $y$, respectively. $NRMSE_y$ is calculated in a test set (which is not used in the estimation) of size $n' = \frac{n}{2}$. For the three cases,  $NRMSE_\beta$ and $NRMSE_y$ decrease as the sample size increases. 
\begin{figure}
    \centering
    \includegraphics[width=150mm]{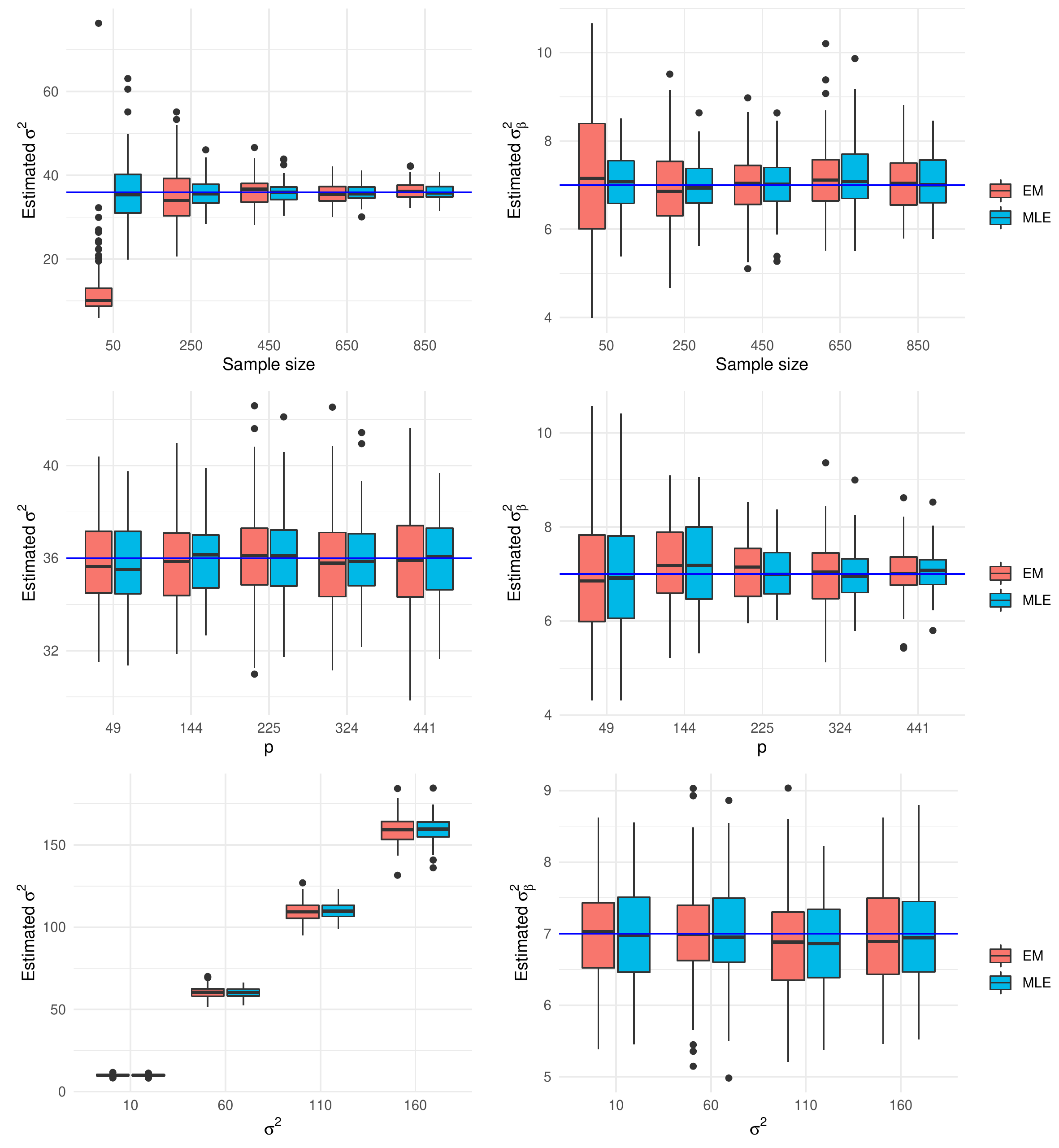}
    \caption{Estimated parameters in the case where the covariance of $\beta$ is diagonal as a function of the sample size, the dimension of $X$, $d$, and the variance $\sigma^2$. Red boxes correspond to EM estimates and the blue ones to MLE estimates. The blue line corresponds to the true value of the parameter $\sigma^2$ and $\sigma_\beta^2$, which are equal to 36 and 7, respectively.}
    \label{fig:fig3}
\end{figure}
\begin{figure}
    \centering
    \includegraphics[width=170mm]{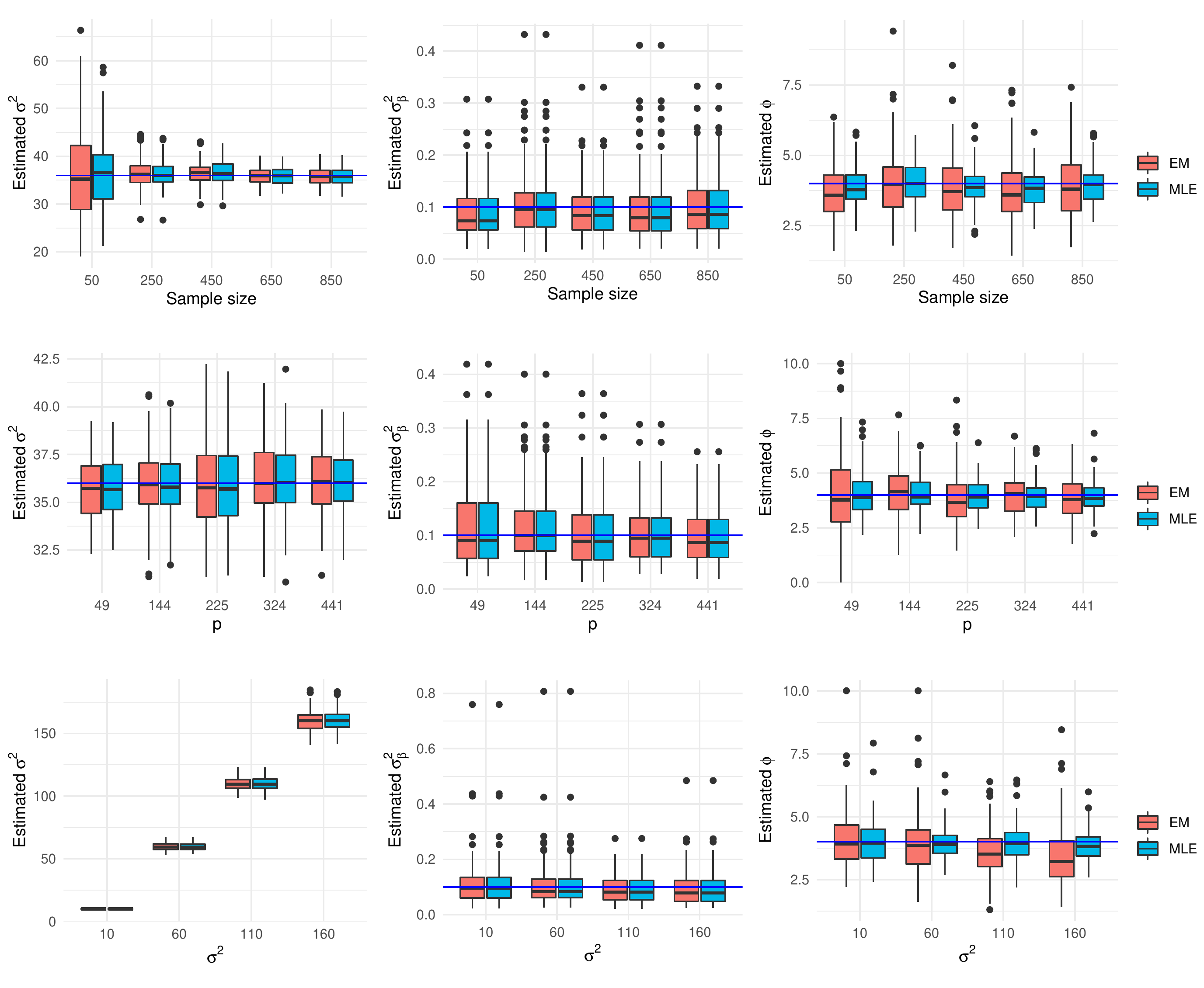}
    \caption{Estimated parameters in the case where the covariance of $\beta$ is the Matérn as a function of the sample size, the dimension of $X$, $d$, and the variance $\sigma^2$. Red boxes correspond to EM estimates and the blue ones to MLE estimates. The blue line corresponds to the true value of the parameter $\sigma^2$, $\sigma_\beta^2$, and $\phi$, which are equal to 36, 0.1, and 4, respectively.}
    \label{fig:fig4}
\end{figure}
\begin{figure}
    \centering
    \includegraphics[width=170mm]{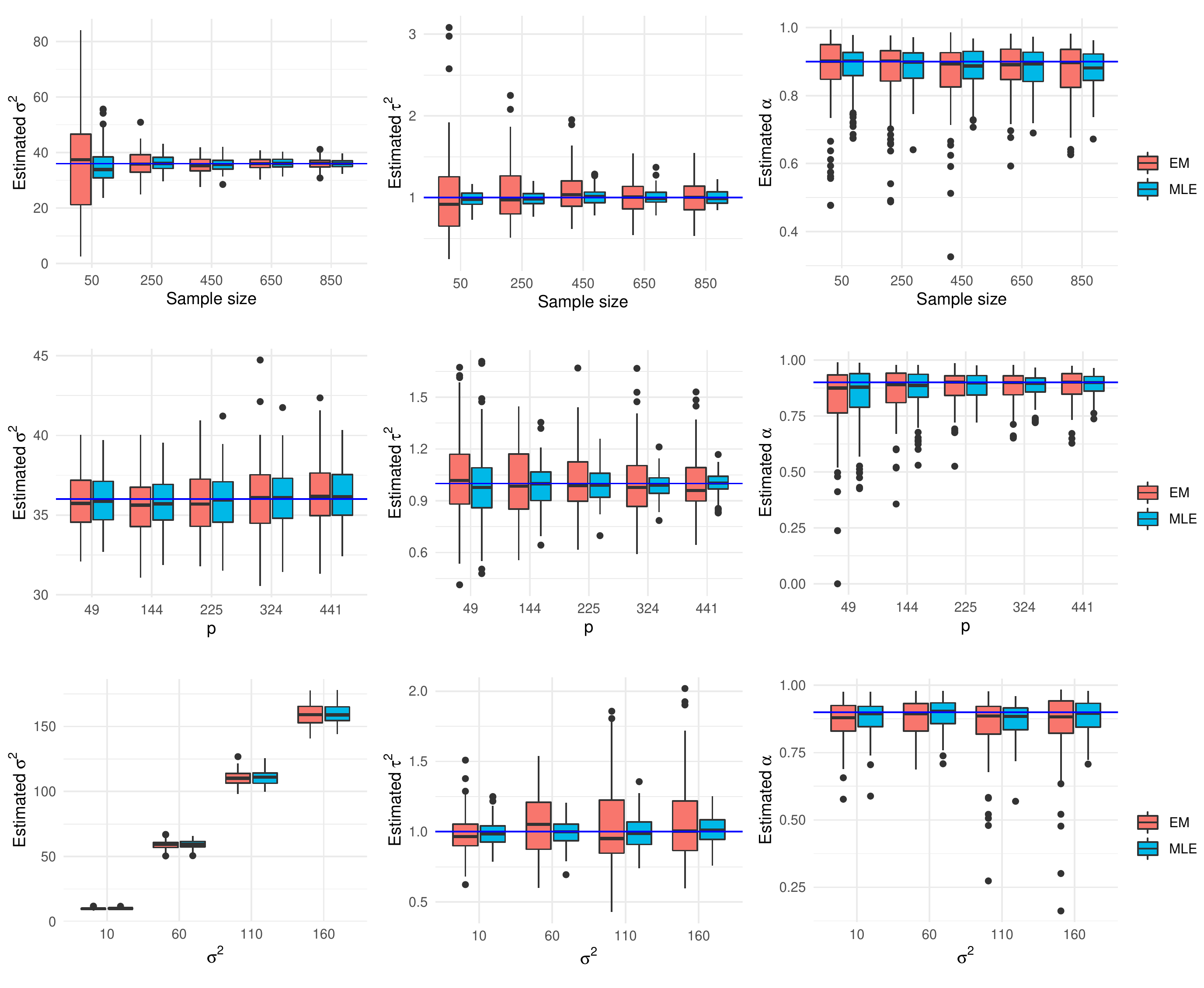}
    \caption{Estimated parameters in the case where the covariance of $\beta$ is the CAR as a function of the sample size, the dimension of $X$, $d$, and the variance $\sigma^2$. Red boxes correspond to EM estimates and the blue ones to MLE estimates. The blue line corresponds to the true value of the parameter $\sigma^2$, $\sigma_\beta^2$, and $\alpha$, which are equal to 36, 1, and 0.9, respectively.}
    \label{fig:fig5}
\end{figure}

To evaluate the parameter estimates, we compare the EM estimates with the maximum likelihood estimates of the parameters, hereafter referred to as MLE, knowing the true $\beta$. More precisely, the MLE estimates are defined as
\begin{equation}
    \Theta_{\text{MLE}} = \arg \max_{\Theta} -\frac{1}{2}\left(\ln(|\Sigma_\theta|)+\beta_{\text{true}}^T\Sigma_\theta^{-1}\beta_{\text{true}} + n\ln(\sigma^2) + \frac{1}{\sigma^2}\lVert y - X\beta_{\text{true}} \rVert^2\right) + C
\end{equation}
where $\beta_{\text{true}}$ is the true $\beta$ simulated for each case with the parameters given in section 3.1. Along with the sample size, we are also interested in how the estimates behave when varying the dimension of $X$, $d$, and the variance parameter $\sigma^2$. Note that in practice, $\Theta_{\text{MLE}}$ cannot be found directly, given that the true $\beta$ is not observed (latent variable). Therefore, we expect the EM algorithm to provide less accurate estimates than MLE. However, we expect that by varying the sample size, the dimension, and the variance $\sigma^2$, the estimations asymptotically will be close to MLE estimates. 

Figures \ref{fig:fig3}, \ref{fig:fig4} and \ref{fig:fig5} show boxplots of EM (red) and MLE (blue) estimates for the diagonal, Matérn and CAR cases as a function of sample size, dimension $d$, and variance $\sigma^2$. For the diagonal case, the estimate of $\sigma^2$ seems to converge to the true value of the parameter (blue line) when the sample size $n$ increases as it does in the usual linear regression model. Note that the estimate of the spatial variance $sigma_\beta^2$ does not seem to converge to the true value of the parameter as the sample size increases, but when $n$ is large enough, EM and MLE seem to provide similar results. This is not unexpected since both methods are based on a single sample of the d-dimensional field $\beta$. As expected, the dimension $d$ also affects the estimate of the parameter $\sigma_\beta^2$, which converges towards the true value as $d$ increases; however, no significant change is observed for $\sigma^2$ when $d$ increases. The effect of the variance $\sigma^2$ on the estimation of $\sigma_\beta^2$ is small, and we observe that for $\sigma^2$ larger than 100, the EM and MLE tend to underestimate $\sigma_\beta^2$.  
Similar behavior can be observed for the Matérn case: the variance parameter $\sigma^2$ seems to converge towards the actual value with increasing sample size. However, there is no significant change in the other parameters (the variance $\sigma_\beta$ and the range $\phi$). The dimension $d$ mainly influences the parameters $\sigma_\beta$ and $\phi$, which describe the spatial structure of the d-dimensional field $\beta$, and as $d$ increases, the estimates converge to the actual values. As for the diagonal case, the EM algorithm underestimates the parameters $\sigma_\beta$ and $\phi$ when the variance $\sigma^2$ increases.
Finally, for the CAR case, the sample size influences the parameters $\sigma^2$ and $\tau^2$, but only slightly the correlation parameter $\alpha$, which is mainly influenced by the dimension $d$. The variance $\sigma^2$ has a significant influence on $\tau^2$, but only a small one on $\alpha$. To summarize:
\begin{itemize}
    \item The sample size $n$ mainly influences the estimation of the variance of the residuals $\sigma^2$ 
    \item The parameters which describe the spatial structure of $\beta$ are mainly influenced by the dimension $d$
    \item As the variance $\sigma^2$ increases, EM underestimates the parameter $\sigma_\beta^2$ of the diagonal and Matérn case, and the range parameter $\phi$
    \item EM estimates are close to MLE estimates in most cases when the sample size and the dimension d are large enough and the variance $\sigma^2$ is small 
\end{itemize}
\begin{figure}
    \centering
    \includegraphics[width=150mm]{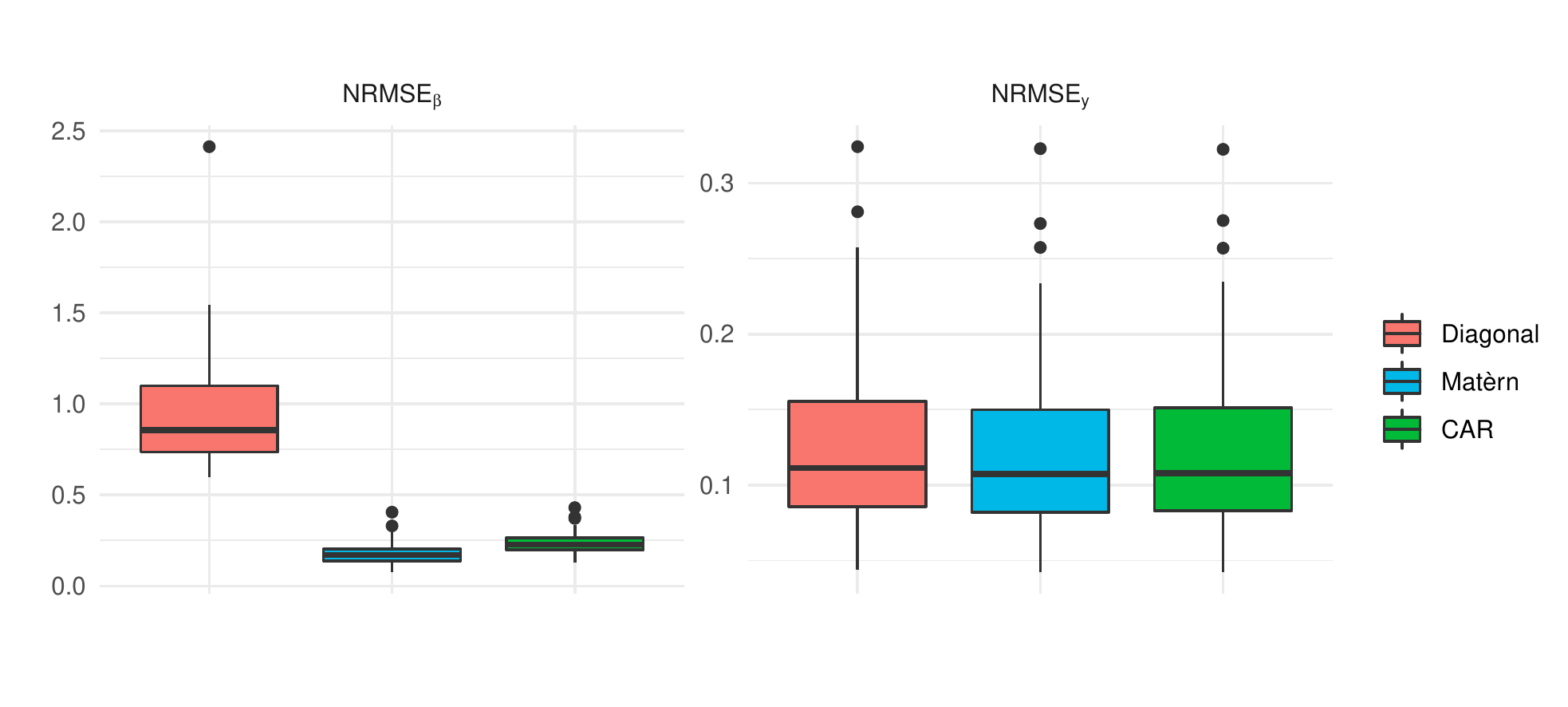}
    \caption{Results of the estimations when the true beta is simulated from Matérn with the parameters $\sigma^2 = 36$, $\sigma_\beta^2 = 0.1$ and $\phi = 4$ and sample size $n=800$. The left panel correspond to $NRMSE_\beta$ and the right one for $NRMSE_y$.}
    \label{fig:fig6}
\end{figure}

Another interesting aspect that needs to be studied is when the coefficients $\beta$ are simulated using one covariance and estimated using another covariance model. To do that, we perform 100 independent simulations of $\beta$ using the Matérn covariance function, and we estimate the parameters using the three cases: diagonal, CAR, and Matérn. Figure \ref{fig:fig6} shows the results of $NRMSE_\beta$ and $NRMSE_y$ of the experiment. It is clear that using the Matérn covariance for the estimation gives better results in terms of $NRMSE_\beta$. Not surprisingly, the diagonal case is the worst model for estimating the coefficients. However, in terms of $NRMSE_y$, there is a small difference between the three methods.  

\section{Application}
\begin{figure}
\centering
\includegraphics[width=110mm]{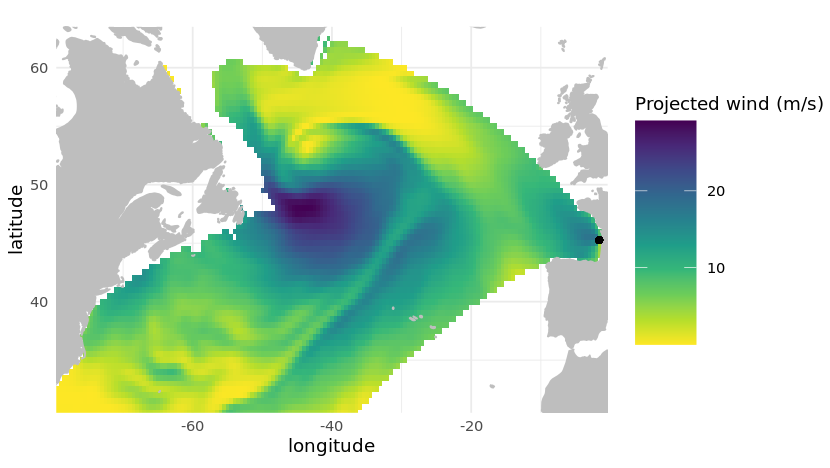}
\caption{CFSR projected wind in the North Atlantic in 1994-01-01 00h:00. The black point represents the target point.}
\label{fig:fig7}
\end{figure}
The proposed method is applied to the problem of predicting the significant wave height ($H_s$) at a location in the Bay of Biscay using wind conditions over the North Atlantic (figure \ref{fig:fig7}), where the significant wave height is the average height of the highest third of the waves, a key measure of wave height that provides information about wave energy.
The data used for $H_s$ comes from the Homere hindcast database \citep{boudiere2013suitable}, and the wind data comes from  Climate
Forecast System Reanalysis (CFSR) \citep{saha2010ncep}. The wind data are pre-processed before being used as a predictor (see \citep{statistical} for the pre-processing procedure). We consider 23 years of $H_s$ and wind data from 1994 to 2016 with a temporal resolution of 3 hours. 

The regression problem is of the form 
\begin{equation}
    H_s(t) = \sum_{j=1}^d X_j(t)\beta_j + \epsilon(t) \,\,\, t=1,...,n
    \label{eq:23}
\end{equation}
where $X_j(t)$ is the predictor at time $t$ and location $j$ defined as
\begin{eqnarray}
\label{eq:eq24}
   &X_j(t;t_j,\alpha_j) = \frac{1}{2\alpha_j+1} \sum_{i = t-t_j-\alpha_j}^{t-t_j+\alpha_j} W_j^2(i),\\ \nonumber
    &t_j + \alpha_j  +1 \leq t \leq t_j -\alpha_j + n
\end{eqnarray}
where $W_j$ is the projected wind (figure \ref{fig:fig7}) defined as 
\begin{equation}
W_j = U_j\,cos\left(\frac{1}{2}(b_j-\theta_j)\right)
\label{eq:eq24}
\end{equation}
$U_j$ is the wind speed, $b_j$ is the great circle bearing, and $\theta_j$ is the wind direction at location $j$. $\alpha_j$ controls the length of the time window, and $t_j$ is the mean travel time of waves which are estimated using the maximum correlation between $H_s$ and the predictor 
\begin{equation}
(\hat{t}_j, \hat{\alpha}_j) = arg\max_{t_j, \alpha_j} \big(corr(H_s,X_j^g(t_j,\alpha_j))\big).
\label{eq:eq9} 
\end{equation}

Let $X = {X_1,...,X_d}$ be the predictor which has the size $67088\times 5651$. Since the predictor has a spatial structure. It is reasonable to assume that the coefficients $\beta$ also have a spatial structure so that nearby locations have close contributions to the waves at the target point. 
This assumption is equivalent to suppose that $\beta \sim \mathcal{N}(0, \Sigma_\theta)$. For the covariance $\Sigma_\theta$, we will consider the cases of Matérn and CAR. For comparison, we also consider the diagonal case even though it does not consider any structure between coefficients. 

 \begin{figure}
    \centering
    \includegraphics[width=160mm]{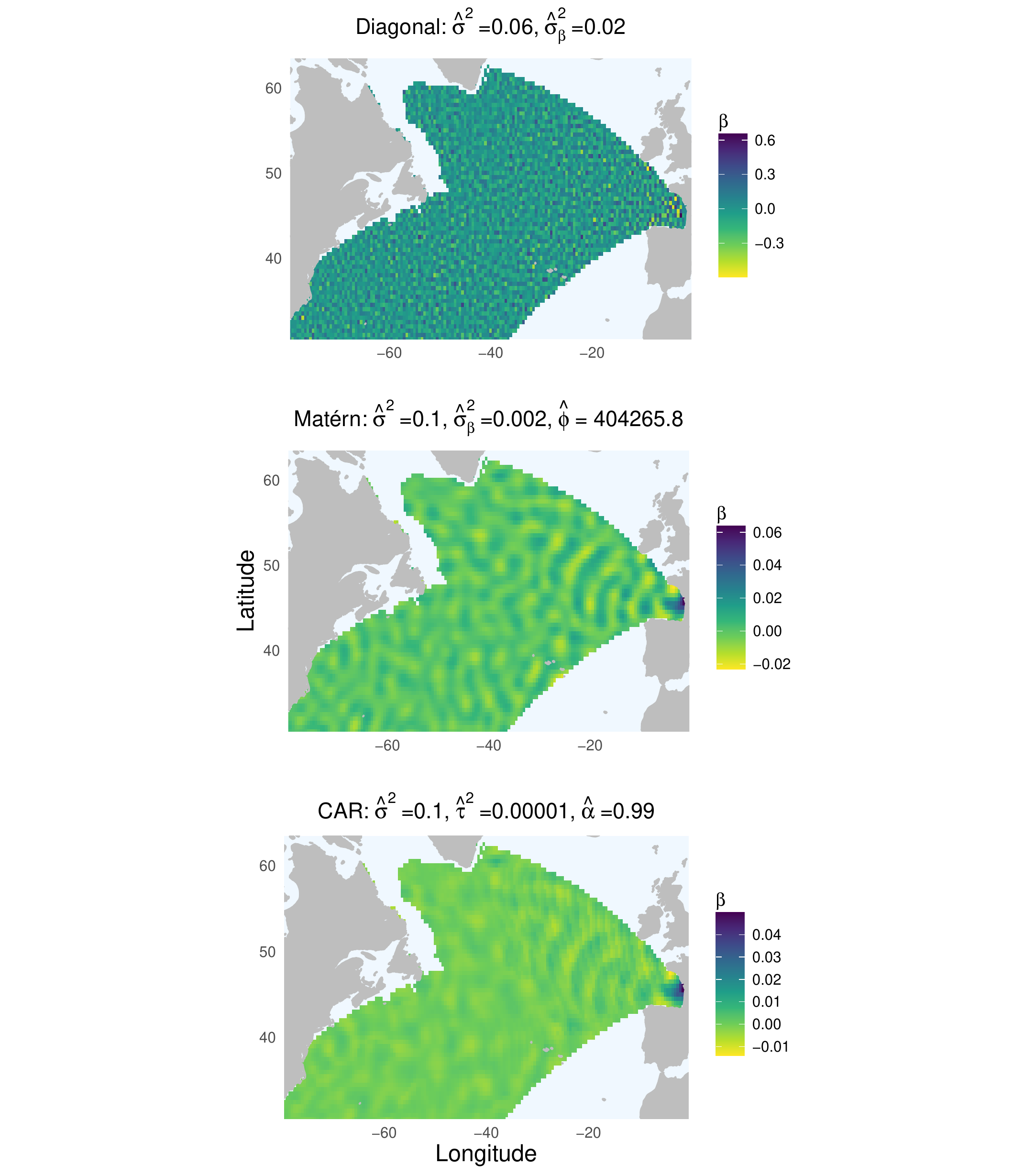}
    \caption{The coefficients $\beta$ estimated using the EM algorithm with diagonal, Matérn, and CAR covariance.}
    \label{fig:fig8}
\end{figure}

The model's parameters (equation \ref{eq:23}) are estimated using data from 1994 to 2013, and the model is evaluated in terms of correlation, RMSE, and bias, using a validation set from 2014 to 2016. Figure \ref{fig:fig8} shows the results of estimating $\beta$ and the covariance parameters using the EM algorithm when the covariance structure is assumed to be diagonal, Matérn and CAR. Not surprisingly, the coefficients estimated with the diagonal covariance show no physical spatial structure. Therefore, the assumption that close locations have close coefficients cannot be taken into account using the diagonal case. This motivates using the Matérn and CAR covariances.    
The Matérn and CAR covariances give the smoothest coefficients with a clear spatial structure. In addition, locations close to the target point have larger coefficients. Therefore, the obtained coefficients are more physically interpretable and take into account our assumption about the covariance. Note that the CAR method is less expensive numerically than the Matérn, which involves inverting the covariance matrix at each iteration of the optimization algorithm used in the M-step.  

\begin{table}
    \centering
    \begin{tabular}{c c c c}
    \hline
    Method & r & RMSE(m)  & bias(m) \\ \hline
    Diagonal & 0.941 & 0.414 &-0.0004\\ 
    Matérn  & 0.956 & 0.354 &-0.04\\ 
    CAR & 0.957 & 0.352 & -0.06 \\ \hline
    \end{tabular}
    \caption{Quantitative comparison of the diagonal, Matérn, and CAR methods in the validation set using the correlation (r), root mean square error (RMSE), and bias.}
    \label{tab:tab1}
\end{table}

Table \ref{tab:tab1} shows the results of the quantitative comparison between the three methods for predicting significant wave height in the validation set using correlation (r), root mean square error (RMSE), and bias. In terms of correlation and RMSE, the diagonal method is the less accurate method. Therefore, adding the spatial structure in the covariance is advantageous in predicting the significant wave height. The CAR and Matérn methods lead to close results regarding r, RMSE, and bias. 
\section{Conclusions}
This study proposed an EM algorithm for estimating generalized Ridge regression with spatial covariates. We have studied three cases: the diagonal, Matérn, and the CAR case. A simulation study is carried out to evaluate the performance of the algorithms, and the EM algorithm successfully estimates the parameters in all cases. We have studied the influence of the sample size, dimension of $X$, and the variance $\sigma^2$ on the estimation. The sample size mainly influences the variance parameter $\sigma^2$. The range parameter of the Matérn and correlation parameter of the CAR are mainly influenced by dimension $d$.

The proposed method is applied to the problem of downscaling the significant wave height in the Bay of Biscay using wind conditions over the North Atlantic. The Matérn method gives smooth coefficients with a clear spatial structure; however, the CAR method slightly outperforms the Matérn method in terms of RMSE. 
The Matérn covariance is clearly a better choice for spatial applications. However, estimating the parameters requires the inversion of the covariance matrix at each iteration of the optimization method in the M-step, which may be a computational bottleneck in many applications. To address this issue, instead of parameterizing the covariance matrix, one can parameterize the precision matrix directly as we did with the CAR method. 
\begin{appendices}
\numberwithin{equation}{section}
\section{Comparison between cross-validation and EM}
 \begin{figure}
    \centering
    \includegraphics[width=150mm]{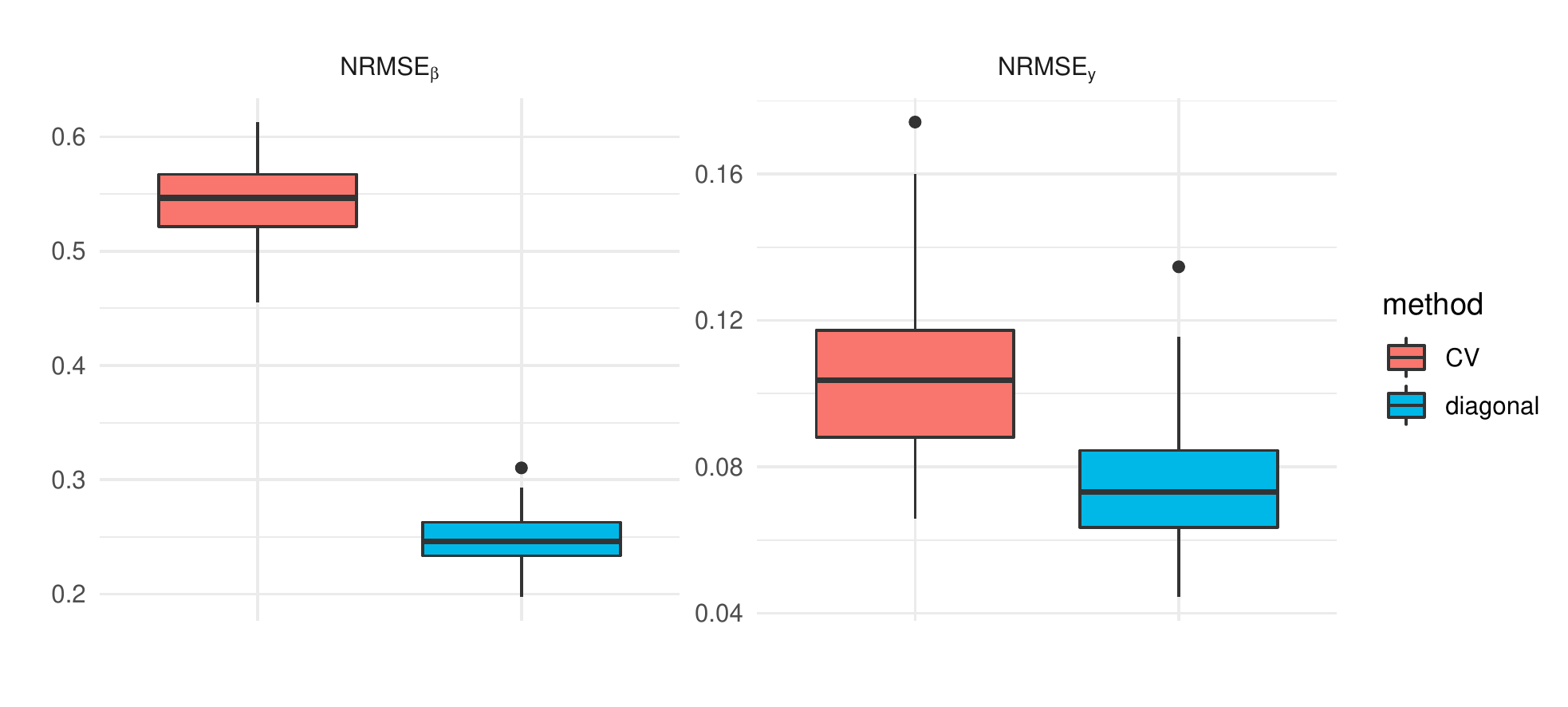}
    \caption{Results of estimating Ridge regression with the EM algorithm and 10-fold cross-validation in the Gaussian case.}
    \label{fig:fig9}
\end{figure}
As stated in section 2, the EM algorithm can be used as an alternative for cross-validation for estimating Ridge regression. In this section, we perform a simulation study to compare the two approaches and use the same simulation procedure discussed in section 3.1. Given the same covariates X (presented in section 3.1) we perform 50 independent random samples of coefficients $\beta$ using the diagonal method (with parameters $\sigma^2 = 36$ and $\sigma_\beta^2 = 7$). For each simulation, we estimate the coefficients using the EM algorithm and the cross-validation method. Figure \ref{fig:fig9} shows the box plot of $NRMSE_\beta$ and $NRMSE_y$. The EM algorithm outperforms cross-validation in estimating the coefficients $\beta$ and predicting $y$.

 \begin{figure}
    \centering
    \includegraphics[width=150mm]{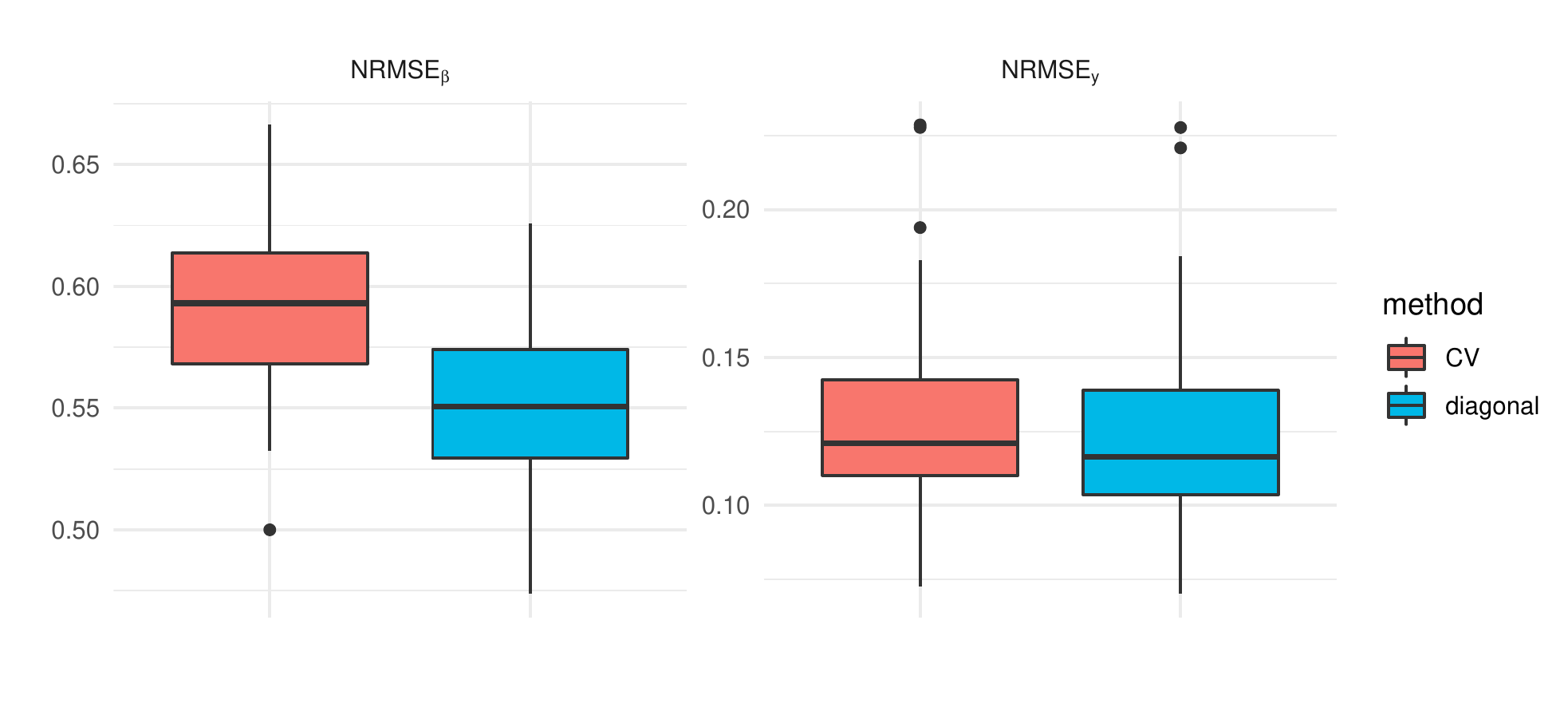}
    \caption{Results of estimating Ridge regression with the EM algorithm and 10-fold cross-validation in the non-Gaussian case.}
    \label{fig:fig10}
\end{figure}
The comparison we performed here is for the Gaussian case; therefore, it is straightforward that the EM algorithm will outperform cross-validation. To see how the two approaches behave in the non-Gaussian case, we simulate the response variable $Y$ using the model
\begin{equation}
    Y = X\beta + \epsilon, \,\,\, \, \text{where} \,\, \epsilon \sim U(2,30)
\end{equation}
Where $U(2,30)$ is the uniform distribution on the interval $[2,30]$.
Figure \ref{fig:fig10} shows the estimation results using the EM algorithm and cross-validation. The EM algorithm still outperforms cross-validation in both $NRMSE_\beta$ and $NRMSE_y$; however, the difference between the two methods here is small than in the Gaussian case.  

\section{The case where $\beta$ has a non-zero mean}
In this section, we consider the case where $\beta$ has a non-zero mean as defined by the hierarchically model  
\begin{equation}
      \begin{split}
            &\beta \sim \mathcal{N}(\mu_\xi,\Sigma_\theta)\\
            &Y \mid \beta, \Theta \sim \mathcal{N}(X\beta, \sigma^2I_n)
      \end{split}
    \label{eq:28}
\end{equation}
where $\Theta = (\sigma^2,\mu_\xi, \theta)$.

The complete log-likelihood is expressed as 
\begin{equation}
\begin{split}
        \ln p(y, \beta;\Theta) &= \ln p(y \mid \beta; \sigma^2)+ \ln p(\beta; \theta) \\
        &= -\frac{1}{2} \left(\ln(|\Sigma_\theta|)+\beta^T\Sigma_\theta^{-1}\beta - 2 \beta^T\Sigma_\theta^{-1}\mu_\xi+ \mu_\xi^T\Sigma_\theta^{-1}\mu_\xi + n\ln(\sigma^2) + \frac{1}{\sigma^2}\lVert y + X\beta \rVert^2 \right) + C
\end{split}
\label{eq:29}
\end{equation}
Where C is a constant. In the M-step, the quantity $Q(\Theta|\Theta^{(t)})$ is maximized with respect to the parameters $\Theta$.\\
\begin{itemize}
    \item E-step: \\
    \begin{equation}
        Q(\Theta|\Theta^{(t)}) = \mathbb{E}(\ln p(y,\beta;\Theta)\mid y,\Theta^{(t)}).
        \label{eq:30}
    \end{equation}
\end{itemize}
The posterior distribution of the latent variable $\beta$ is a normal distribution with mean $\mu_{\beta \mid y}$ and covariance matrix $\Sigma_{\beta \mid y}$ such that 
\begin{equation}
\left\{
    \begin{array}{ll}
        \Sigma_{\beta|y} = 
        (\Sigma_\theta^{-1} + \frac{1}{\sigma^2}X^TX)^{-1}\\
        \mu_{\beta|y} = \Sigma_{\beta|y}(\Sigma_\theta^{-1}\mu_\xi + \frac{1}{\sigma^2}X^Ty).
    \end{array}
\right.
\label{eq:31}
\end{equation}
Therefore, 
\begin{equation}
     Q(\Theta|\Theta^{(t)}) = -\frac{1}{2}\left(\ln(|\Sigma_\theta|) + \mathrm{Tr}(\Sigma_\theta^{-1}\mathbb{E}(\beta\beta^T\mid y,\Theta^{(t)}))- 2 \mu_{\beta|y}^T\Sigma_\theta^{-1}\mu_\xi+ \mu_\xi^T\Sigma_\theta^{-1}\mu_\xi+ n\ln(\sigma^2)+\frac{1}{\sigma^2}\mathbb{E}(\lVert y-X\beta\rVert^2\mid y,\Theta^{(t)})\right) + C
     \label{eq:33}
\end{equation}
where 
\begin{equation}
    \left\{
    \begin{array}{ll}
        \mathbb{E}(\beta\beta^T|y;\Theta^{(t)}) = \Sigma_{\beta|y} + \mu_{\beta|y}\mu_{\beta|y}^T\\
        \mathbb{E}(\lVert y - X\beta\rVert^2|y;\Theta^{(t)}) = \lVert y\rVert^2 -2y^TX\mu_{\beta|y}+
        \mathrm{Tr}(X^TX\mathbb{E}(\beta\beta^T|y;\Theta^{(t)}))
    \end{array}
\right.
\label{eq:34}
\end{equation}
\begin{itemize}
    \item M-step:
\end{itemize}
The maximization step computes
\begin{equation}
    \Theta^{(t+1)} = \arg \max_{\Theta}  Q(\Theta|\Theta^{(t)})
    \label{eq:35}
\end{equation}
which leads to the following updates of the parameters
\begin{equation}
\begin{split}
    &\sigma^{2,(t+1)} = \frac{1}{n}(\lVert y\rVert^2 -2y^TX\mu_{\beta|y}+
       \mathrm{Tr}(X^TX\mathbb{E}(\beta\beta^T|y;\Theta^{(t)})))\\
    & (\xi^{(t+1)},\theta^{(t+1)}) = \arg \max_{\xi,\theta} \, \, \ln(|\Sigma_\theta^{-1}|) - \mathrm{Tr}(\Sigma_\theta^{-1}\mathbb{E}(\beta\beta^T\mid y,\Theta^{(t)})) + 2 \mu_{\beta|y}^T\Sigma_\theta^{-1}\mu_\xi^{(t)}- \mu_{{\xi}^{(t)}}^T\Sigma_\theta^{-1}\mu_\xi^{(t)}
\end{split}
\label{eq:36}
\end{equation}
\end{appendices}
\bibliographystyle{cas-model2-names}
\bibliography{sreg}

\end{document}